\documentclass[aps,pra,twocolumn,superscriptaddress,showkeys]{revtex4-1}

\usepackage{graphicx,dcolumn,hyperref,braket,amsmath,amssymb} 
\usepackage{epstopdf,color}
\usepackage{natbib}
\usepackage{textcomp}

\begin{document}

\title{Hyperradiance from collective behavior of coherently driven atoms}

\author{M.-O. Pleinert}
%\thanks{Correspondence and requests for materials should be addressed to M.-O.P. (marc.pleinert@fau.de)}
\affiliation{Institut f\"{u}r Optik, Information und Photonik, Friedrich-Alexander-Universit\"{a}t Erlangen-N\"{u}rnberg (FAU), 91058 Erlangen, Germany}
\affiliation{Department of Physics, Oklahoma State University, Stillwater, Oklahoma 74078, USA}
\affiliation{Erlangen Graduate School in Advanced Optical Technologies (SAOT), Friedrich-Alexander-Universit\"{a}t Erlangen-N\"{u}rnberg (FAU), 91052 Erlangen, Germany}
\author{J. von Zanthier}
\affiliation{Institut f\"{u}r Optik, Information und Photonik, Friedrich-Alexander-Universit\"{a}t Erlangen-N\"{u}rnberg (FAU), 91058 Erlangen, Germany}
\affiliation{Erlangen Graduate School in Advanced Optical Technologies (SAOT), Friedrich-Alexander-Universit\"{a}t Erlangen-N\"{u}rnberg (FAU), 91052 Erlangen, Germany}
\author{G. S. Agarwal}
\affiliation{Department of Physics, Oklahoma State University, Stillwater, Oklahoma 74078, USA}
\affiliation{Institute for Quantum Science and Engineering and Department of Biological and Agricultural Engineering, Texas A\&M University, College Station, Texas 77843, USA}

%\date{\today}

\begin{abstract}
The collective behavior of ensembles of atoms has been studied in-depth since the seminal paper of Dicke [R. H. Dicke, Phys. Rev. \textbf{93}, 99 (1954)], where he demonstrated that a group of emitters in collective states is able to radiate with increased intensity and modified decay rates in particular directions, a phenomenon which he called superradiance. Here, we show that the fundamental setup of two atoms coupled to a single-mode cavity can be distinctly exceeding the free-space superradiant behavior, a phenomenon which we call hyperradiance. The effect is accompanied by strong quantum fluctuations and surprisingly arises for atoms radiating out-of-phase, an alleged non-ideal condition, where one expects subradiance. We are able to explain the onset of hyperradiance in a transparent way by a photon cascade taking place among manifolds of Dicke states with different photon numbers under particular out-of-phase coupling conditions. The theoretical results can be realized with current technology and thus should stimulate future experiments.
\end{abstract}

\maketitle

\section{Introduction}

Arguably one of the most enigmatic phenomena in the history of quantum optics is the discovery of superradiance by Dicke \cite{Dicke:1954,Rehler:1971,Bonifacio:1971,Friedberg:1973,Agarwal:1974,Gross:1982,Skribanowitz:1973,Vrehen:1977,Kaiser:2016,Thiel:2007,Maser:2009,Blatt:2008,Monz:2011,Wiegner:2011,Scully:2009,Scully:2009a,Svidzinsky:2013,Bienaime:2013,Feng:2014,Scully:2015,Longo:2016,Svidzinsky:2016,Rohlsberger:2010,Rohlsberger:2013,Bhatti:2015,DeVoe:1996}. 
A key requirement in Dicke's work is the initial preparation of an ensemble of two-level atoms in a special class of collective states, so-called symmetric Dicke states. The startling gist is that even though atoms in these states have no dipole moment, they radiate with an intensity which is enhanced by a factor of $N$ compared to $N$ independent atoms.
The preparation of such states has been a challenge since.
The first experiments on superradiance were done in atomic vapors \cite{Skribanowitz:1973,Vrehen:1977,Gross:1982}, where it was assumed that the fully excited system in the course of temporal evolution would at some time be found in a Dicke state leading to the emission of superradiant light \cite{Rehler:1971,Bonifacio:1971,Gross:1982}. 
The same assumption led to the recent observation of subradiance \cite{Kaiser:2016}.

\begin{figure*}
	\centering \includegraphics{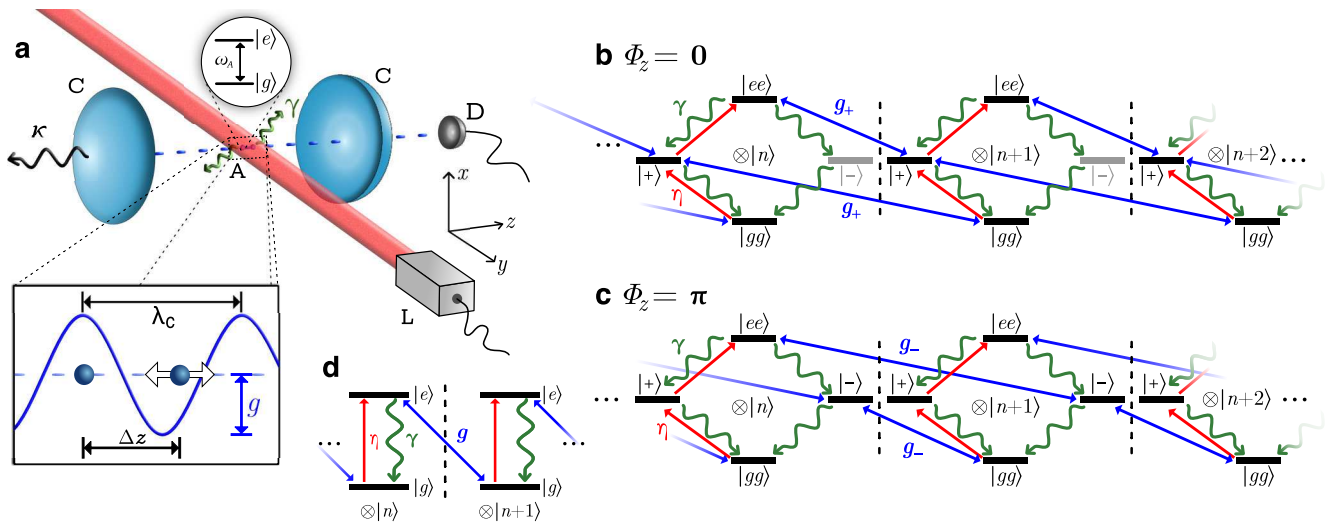}
	\caption{\label{fig:cavity_setup} Basic properties of the system. (a) Sketch of the system consisting of two atoms (A) that are coupled to a single-mode cavity (C) and driven by a coherent laser (L) with Rabi frequency $\eta$. Intracavity photons can leak through the mirrors by cavity decay ($\kappa$) and be registered by a detector (D). Another possible dissipative process is spontaneous emission ($\gamma$) by the atoms. The inset shows a magnified section of the arrangements of the atoms: One atom (depicted left) is fixed at an anti-node of the cavity field, while the other atom (right) can be scanned along the cavity axis causing a relative phase shift $\phi_z$ between the radiation of the atoms. (b,c) Energy levels and transitions of the system for (b) in-phase and (c) out-of-phase radiation of the atoms. In the case of two atoms, the state space consists of manifolds of four Dicke states with different intracavity photon numbers: For a fixed cavity state $\ket{n}$, the unentangled two atom ground and two atom excited state $\ket{gg}$ and $\ket{ee}$, respectively, as well as the maximally entangled symmetric and anti-symmetric Dicke states $\ket{\pm}$. For clarity, we neither draw the transitions due to cavity decay nor the detuning. (d) Energy levels and transitions of the corresponding system containing a single atom.}
\end{figure*}

Many of the mysteries behind this effect have begun to unfold only recently. 
In the seventies, it was realized that the superradiant emission results from the strong quantum correlations among the atoms being prepared in symmetric Dicke states \cite{Agarwal:1974,Gross:1982}. Only lately, it became clear that the $N$-fold radiative enhancement can be explained by the multiparticle entanglement of the Dicke states. 
For example, studying superradiance in a chain of Dicke-entangled atoms on a lattice enables one to identify that multiple interfering quantum paths lead to the collective subradiant and superradiant behavior \cite{Nienhuis:1987,Wiegner:2011}. 
The strong entanglement of the states can already be inferred from the simpler two atom case and holds for almost all Dicke states of a multi-atom-system. 
With the current advances in quantum information science, we indeed understand the great difficulties in precisely preparing such highly entangled multiparticle states.
There are proposals for the generation of whole classes of Dicke states using projective measurements \cite{Thiel:2007,Bastin:2009,Maser:2009}, yet these schemes have very low success probability. Deterministic entanglement has been produced with about a dozen qubits in the form of $W$-states \cite{Blatt:2008,Monz:2011}, but we are still far away from the realization of $W$-states for an arbitrary number of qubits.
Calculations show that these states, which can be considered to be the analog to single excitation Dicke states with appropriate phase factors, produce also enhancement by a factor of $N$ \cite{Scully:2009,Scully:2009a,Wiegner:2011}.  

In comparably simpler systems, yet with a higher number of excitations, one can also study the quantum statistical aspects of the collective emission \cite{Bhatti:2015}.
However, down to the present day the generation and measurement of higher excited multi-particle entangled Dicke states are challenging, so mostly systems with no more than one excitation have been realized. In these systems, the dynamics is still quite complex, yet superradiance and also subradiance can be fruitfully explored \cite{Scully:2009,Scully:2009a,Svidzinsky:2013,Feng:2014,Scully:2015,Longo:2016,Svidzinsky:2016,Bienaime:2013,Wiegner:2011}. 
Experiments on superradiance with single photon excited Dicke states were also reported for nuclear transitions \cite{Rohlsberger:2010,Rohlsberger:2013}. A recent work discusses preparation of a single photon subradiant state and its radiation characteristics for atoms in free space \cite{Scully:2015}. Even applications of superradiance are beginning to appear. Lately, a laser with a frequency linewidth less than that of a single-particle decoherence linewidth was realized \cite{Bohnet:2012} by using more than one million intracavity atoms and operating in a steady-state superradiant regime \cite{Meiser:2010a,Meiser:2010}.

Despite these advances, one is still faced with the difficulties in the optical domain which arise from the infinite number of modes in free space and interatomic effects like the dipole-dipole interaction \cite{Agarwal:1974,Friedberg:1973}. It is thus evident that one needs to work with systems which have fewer degrees of freedoms and where a precise preparation of entangled states is possible. This brings us to work with single-mode cavities \cite{DeVoe:1996,DeVoe:1984} with few atoms. The current technological progress in atom trapping and the availability of well characterized single-mode cavities is making this ideal situation becoming more and more a reality. Several experiments in the last two years have been reported using such well-characterized systems \cite{Reimann:2015,Casabone:2015,Neuzner:2016}.  
The experiments consist of two coherently driven atoms \cite{Reimann:2015,Neuzner:2016} or entangled ions \cite{Casabone:2015} coupled to a single-mode cavity as depicted in Fig. \ref{fig:cavity_setup}(a). This setup enables one to study collective behavior as a function of various atomic and cavity parameters, e.g., the precise location of the atoms. 

In spite of this recent progress on superradiant and subradiant behavior \cite{Scully:2009,Scully:2009a,Svidzinsky:2013,Feng:2014,Scully:2015,Longo:2016,Svidzinsky:2016,Bienaime:2013,Wiegner:2011} and the surge of new classes of experiments \cite{Reimann:2015,Casabone:2015,Neuzner:2016,Bohnet:2012}, there is yet no report of atomic light emission beyond that of superradiance. 
In this paper, we demonstrate that a two-atom system coupled to a single-mode cavity is capable of radiating up to several orders of magnitude higher than a corresponding system consisting of two uncorrelated atoms, thereby exceeding the free-space superradiant emission by far. We call this effect hyperradiance. Surprisingly, hyperradiance occurs in a regime which one usually considers to be non-ideal, namely when the two atoms radiate out-of-phase. Such nonideal conditions are rather expected to suppress superradiance and thus one is not inclined to imagine in this regime an emission burst exceeding the one of superradiance. 

Although the study that we present is in the context of atomic systems, the results should be applicable to other types of two-level systems like ions \cite{DeVoe:1996,Stute:2012,Casabone:2015}, superconducting qubits \cite{Wallraff:2004,Fink:2009,Mlynek:2014} and quantum dots \cite{Hennessy:2007,Faraon:2008,Miguel-Sanchez:2013,Rundquist:2014,Leymann:2015}. We thus expect that our findings stimulate a multitude of new experiments in various domains of physics.

\section{Methods}

\subsection{System}
The investigated system follows the experiments of \cite{Reimann:2015,Neuzner:2016} consisting of two atoms (A) coupled to a single-mode cavity (C) as shown in Fig. \ref{fig:cavity_setup}(a). A laser (L) oriented perpendicular to the cavity axis coherently drives the atoms. In this paper, we fix one atom at an antinode of the cavity field, while we vary the position of the other atom along the cavity axis inducing a relative phase shift between the radiation of the atoms. The atoms within the cavity are modeled as two-level systems with transition frequency $\omega_A$, driven by a laser field at frequency $\omega_L$, and couple to a single-mode of the cavity with frequency $\omega_C=2\pi c/\lambda_C$. 
The $i$th atom is characterized by spin-half operators $S_i^+ = \ket{e}_i\bra{g}_i$, $S_i^- = (S_i^+)^\dagger$ and $S_i^z = (\ket{e}_i\bra{e}_i - \ket{g}_i\bra{g}_i)/2$. 
Bosonic annihilation and creation operator $a$ and $a^\dagger$ describe the intracavity mode. 
The dynamical behavior of the entire system can be treated in a master equation approach \cite{Agarwal:2012} and is governed by 
\begin{equation}
\label{eq:master}
\frac{d}{dt} \rho =  -\frac{i}{\hbar} \left[H_0 + H_I + H_L, \rho \right] + \mathcal{L}_\gamma \rho + \mathcal{L}_\kappa \rho \, ,
\end{equation}
where $\rho$ is the density operator of the atom-cavity system.
In the interaction frame rotating at the laser frequency, atoms and cavity are described by $H_0 = \hbar \Delta (S^z_1+S^z_2) + \hbar \delta a^\dagger a$. Here, $\Delta=\omega_A-\omega_L$ is the atom-laser detuning and $\delta=\omega_C-\omega_L$ the cavity-laser detuning.
The Tavis-Cummings interaction term of atom-cavity coupling is given by 
$H_I = \hbar \sum_{i=1,2} g_i \left( S_i^+ a + S_i^- a^\dagger \right)$
and can be obtained by utilizing the dipole approximation and applying the rotating wave approximation \cite{Tavis:1968}.
The term $g_i=g\cos (2\pi z_i / \lambda_C)$ describes the position-dependent coupling strength between cavity and $i$th atom. The interatomic distance $\Delta z$ induces a phase shift $\phi_z=2\pi\Delta z / \lambda_C$ between the radiation emitted by the two atoms. Since $\phi_z$ can be chosen $\pmod {2\pi}$, separations of the atoms much larger than the cavity wavelength $\lambda_C$ can be achieved in order to avoid direct atom-atom interactions as in \cite{Goldstein:1997}. Observe that in our setup at $\phi_z=\pi/2 \, (3\pi/2)$, only one atom is coupled to the cavity. 
The coherent pumping of the atoms is characterized by the Hamiltonian $H_L = \hbar \eta \sum_{i=1,2} \left( S_i^+ + S_i^-  \right)$. Hereby, it is assumed that the pumping laser with Rabi frequency $\eta$ propagates perpendicular to the cavity axis. Neglecting possible interatomic displacements in $y$-direction leads to a homogeneous driving of the atoms. Varying pump rates due to spatial variation of the laser phase could be absorbed into effective coupling constants of the atoms \cite{Casabone:2015}.
For fixed atomic transition dipole moment, $\eta$ indicates the strength of the coherent pump.
Spontaneous emission of the atoms at rate $\gamma$ is taken into account by the term $\mathcal{L}_\gamma \rho = \gamma/2 \sum_{i=1,2} \left( 2 S_i^- \rho S_i^+ - S_i^+S_i^- \rho - \rho S_i^+ S_i^- \right)$, whereas cavity decay at rate $\kappa$ is considered by the Liouvillian $\mathcal{L}_\kappa \rho =  \kappa/2 \left( 2a\rho a^\dagger -a^\dagger a \rho - \rho a^\dagger a \right)$.
In this letter, we neglect marginal dephasing effects, which, for example, become relevant in the case of quantum dots.

In order to work out the dynamical behavior of the atom-cavity system, we have to solve Eq. \eqref{eq:master}, which depends on many parameters. Whereas $\eta$, $\delta$, and $\Delta$ can be easily varied, $g$, $\kappa$, and $\gamma$ are intrinsic properties and depend on the design of the cavity and the atomic system used. 
The specific dynamics very much depends on the cavity coupling and the cavity $Q$-factor. Thus to keep our discussion fairly general it becomes necessary to solve the master equation quite universally so that the behavior in different regimes can be studied. We thus resort to numerical techniques based on QuTiP \cite{Johansson:2012}. We ensured the numerical convergence of our results by considering different cutoffs of the photonic Hilbert space.
%In this letter, we investigate the radiance mainly with respect to the freely adjustable parameters $\eta$ and $\phi_z$ for different atom-cavity-laser setups.

\subsection{Transitions}
To clarify the dynamical behavior of the system, we make use of the collective basis states $\ket{gg}$, $\ket{ee}$ and $\ket{\pm}$ to describe the atoms. The symmetric and anti-symmetric Dicke state $\ket{\pm}=D_\pm^\dagger \ket{gg}=(\ket{eg} \pm \ket{ge})/\sqrt{2}$ are created by the collective Dicke operators $D_\pm^\dagger=(S_1^+ \pm S_2^+)/\sqrt{2}$ \cite{Agarwal:2012}. 
Rewriting interaction and pumping Hamiltonian in terms of the collective operators $D_\pm^\dagger$ \cite{Fernandez-Vidal:2007}, yields a clear picture of the occurring transitions as can be seen in Fig. \ref{fig:cavity_setup}(b) and (c). The pumping term is then given by $H_L=\hbar \sqrt{2}\eta(D_+^\dagger + D_+)$ and gives rise to the transitions $\ket{gg,n}\overset{\eta}{\rightarrow}\ket{+,n}\overset{\eta}{\rightarrow}\ket{ee,n}$ with $n$ being the number of photons in the cavity mode. Hence, only symmetric Dicke state $\ket{+}$ and doubly excited state $\ket{ee}$ are pumped. 
The interaction term, on the other hand, couples the cavity to $\ket{+}$ or $\ket{-}$ depending on the interatomic phase $\phi_z$. It reads $H_I=H_+ + H_-$ with $H_\pm = \hbar g_\pm(\phi_z) (a D_\pm^\dagger+a^\dagger D_\pm)$ and $g_\pm(\phi_z)=g(1\pm \cos (\phi_z)) /\sqrt{2}$.

In case of an in-phase radiation of the atoms, apparently $g_{-}(\phi_z=0)=0$ and the anti-symmetric Dicke state $\ket{-}$ is uncoupled from the dynamics. Possible atom-cavity interactions are then via the states $\ket{ee,n}\overset{g_{+}}{\longleftrightarrow}\ket{+,n+1}\overset{g_{+}}{\longleftrightarrow}\ket{gg,n+2}$, see also Fig. \ref{fig:cavity_setup}(b).

For atoms radiating out of phase, however, $g_+(\phi_z=\pi)=0$ and the cavity only couples via $\ket{-}$, i.e. $\ket{ee,n}\overset{g_{-}}{\longleftrightarrow}\ket{-,n+1}\overset{g_{-}}{\longleftrightarrow}\ket{gg,n+2}$. Note that although only the symmetric Dicke state $\ket{+}$ is pumped by the applied coherent field, the photon number in the cavity is non-zero for an out of phase radiation of the atoms due to higher-order processes, which can populate the state $\ket{-}$. 
These are direct cavity coupling $\ket{ee,n} \overset{g_{-}}{\rightarrow} \ket{-,n+1}$ and spontaneous emission $\ket{ee,n} \overset{\gamma}{\rightarrow} \ket{\pm,n} \overset{\gamma}{\rightarrow} \ket{gg,n}$, see also Fig. \ref{fig:cavity_setup}(c). Note that the latter process, of course, takes place for $\phi_z=0$ as well as $\phi_z=\pi$. 

For a phase in between, both couplings are present as $g_-(\phi_z)$ and $g_+(\phi_z)$ will be nonzero. For the sake of completeness, we list the transitions due to cavity decay which read $\ket{.,n} \overset{\kappa}{\rightarrow} \ket{.,n-1}$ and are possible for all values of $\phi_z$. 

\begin{figure*}
	\centering \includegraphics{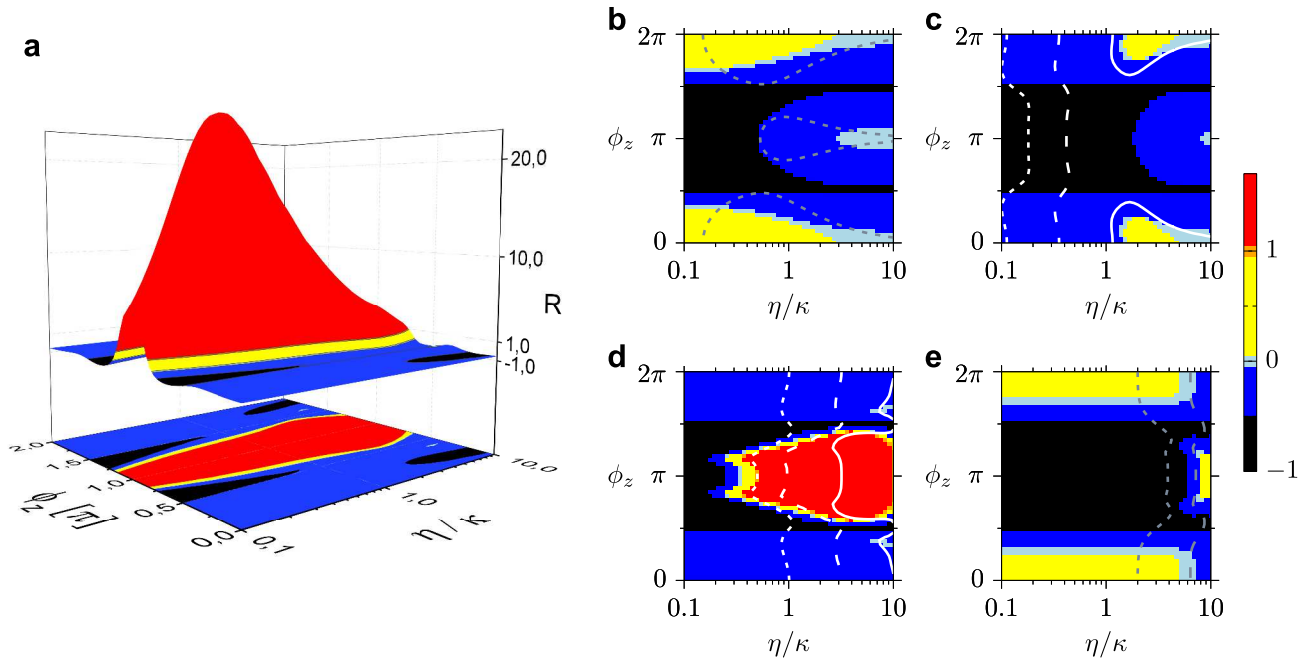}
	\caption{\label{fig:R_outofphase} Radiance witness $R$ for different regimes as a function of the interatomic phase $\phi_z$ and pumping rate $\eta$. The color encodes six different regimes of radiation, i.e., extremely subradiant (black), subradiant (blue), uncorrelated (light blue), enhanced (yellow), superradiant (orange), and hyperradiant (red). Dotted, dashed, and solid curve in the figures indicate the mean photon numbers $\braket{a^\dagger a}_2=0.01,0.1,1$, respectively.
		(a) 3D plot and 2D surface map of the predominant hyperradiant area for $\gamma=\kappa$, $g=10\kappa$ and no detuning. Here, the superradiant and uncorrelated scattering area are very small and can hardly be seen.	
		(b,c) Results for bad and intermediate cavity with $\gamma=\kappa$, no detuning and (b) $g=0.1\kappa$, (c) $g=\kappa$.
		(d,e) Influence of the detuning on hyperradiance with $\gamma=\kappa$, $g=10\kappa$ and (d) $\delta=\Delta=\kappa$, (e) $\delta=\Delta=10\kappa$.}
\end{figure*}

\subsection{Radiance witness $R$}

%Typically, superradiance effects are investigated with atomic correlation operators like $\braket{S^+S^-}$. 
In the considered setup, it is natural to measure the emitted radiation at an external detector (D) placed along the cavity axis, see Fig. \ref{fig:cavity_setup}(a). As the pumping beam (L) is perpendicular to the cavity axis and thus will not contribute photons along the cavity axis, the registered mean photon number at the detector (D) is proportional to the corresponding intracavity quantity.
By performing a reference simulation of a single atom located at an antinode of the cavity field, we are thus able to quantify the radiant character of the two-atom system as a function of the correlations of the two atoms by use of a radiance witness 
\begin{equation}\label{eq:superradiance-witness}
R:=\frac{\braket{a^\dagger a}_{2}-2 \braket{a^\dagger a}_{1}}{2 \braket{a^\dagger a}_{1}} \, ,
\end{equation}
involving the intracavity bosonic operators $a$ and $a^\dagger$.
Here, $\braket{a^\dagger a}_i$ is the steady-state mean photon number with $i=1,2$ atoms in the cavity. The factor $2$ arising in front of $\braket{a^\dagger a}_{1}$ results from the comparison of the coupled two-atom system to the system of two uncorrelated atoms, while the denominator in Eq. \eqref{eq:superradiance-witness} yields a normalization of $R$.

The witness $R$ is composed of experimental observables, i.e., number of photons, which can be measured as in \cite{Reimann:2015}. A possible detection strategy for $R$ is, for instance, scanning the second atom from $\phi_z=\pi/2$ to $\phi_z=\theta$ simulating the transition from effectively one atom coupled to the cavity to two atoms radiating in-phase ($\theta=0$) or out-of-phase ($\theta=\pi$)  into the cavity mode. By evaluating the experimental data according to Eq. \eqref{eq:superradiance-witness}, the radiance witness $R$ can be obtained.

$R=0$ reveals an uncorrelated scattering, where the scattering of the two atom-cavity system is simply the sum of two independent atoms in the cavity. 
A value of $R$ different from zero thus indicates correlations between the atoms.
Negative or positive values of $R$ signal a suppressed or enhanced radiation of the two atom-cavity system, respectively. 
$R=1$, in particular, implies that the radiation scales with the square of the number of atoms $\propto N^2$, which is called superradiance with respect to the free-space scenario \cite{Dicke:1954}.
Atoms confined to a cavity, however, feel a back-action of the cavity field which modifies their collective radiative behavior, allowing for a remarkably new possibility $R>1$. In fact, we found regimes with $\braket{a^\dagger a}_2 > 50 \braket{a^\dagger a}_1$ yielding $R$ greater than $24$. In order to emphasize this phenomenon, we call the domain of $R>1$ hyperradiant.

The atomic correlation quantity $\braket{S^+S^-}$, on the other hand, can be used to obtain the sideway radiation of the atoms. Note that in the bad cavity regime, $R$ reduces to the definition in terms of atomic operators like in \cite{Meiser:2010a,Meiser:2010,Bohnet:2012} due to adiabatic elimination, while in good cavities with $g > \gamma$, the emission of photons into the cavity mode dominates over spontaneous emission in side-modes. $R$ thus constitutes a very natural witness for the setup of Fig. \ref{fig:cavity_setup}(a).

\subsection{Semiclassical treatment}
Several phenomena of fundamental atom-light interaction can be fully analyzed within a semiclassical framework, even atoms coupled to a cavity in the weak atomic excitation limit. In a semiclassical approximation, one decouples the dynamics of atoms and cavity, i.e., $\braket{aS^z} \approx \braket{a}\braket{S^z}$ and assumes a vanishing atomic excitation leading to $\braket{S_i^z}\approx -1/2$. In steady state, one is able to deduce an analytical result for $\braket{a}$, which is proportional to the intracavity field. In terms of the parameters of the system, it reads
\begin{equation}\label{eq:classical-field}
\braket{a}=\frac{\eta}{g} \frac{N\mathcal{G}}{\frac{1}{g^2}\left(\frac{\gamma}{2}+i\Delta\right)\left(\frac{\kappa}{2}+i\delta\right)-N\mathcal{H}} \, ,
\end{equation}
where $N$ is the number of atoms inside the cavity. We further introduced the two collective coupling parameters, $\mathcal{H}=N^{-1}\sum_{i=1}^N \cos^2 [2\pi z_i/\lambda_C]$ along the cavity and $\mathcal{G}=N^{-1}\sum_{i=1}^N \cos [2\pi z_i/\lambda_C]$ for the incident beam \cite{Tanji-Suzuki:2011}, which involve the position-dependent atom-cavity couplings $g_i$.
In the investigated two-atom system, these can be written as a function of the interatomic phase only: $\mathcal{H}(\phi_z)=[1+\cos^2(\phi_z)]/2$ and $\mathcal{G}(\phi_z)=[1+\cos(\phi_z)]/2$. Equation \eqref{eq:classical-field} can also be derived in a classical framework, where the atoms are treated as radiating dipoles that couple to a non-quantized standing-wave optical resonator \cite{Tanji-Suzuki:2011}. Hereby, one exploits the condition that the intracavity field needs to match itself after a round trip in order to be sustained by the resonator.
Observe that the \mbox{(semi-)}classical intracavity field is proportional to $\mathcal{G}(\phi_z)$. For an out-of-phase configuration, it holds $\mathcal{G}(\phi_z=\pi)=0$ and thus semiclassical treatment predicts a vanishing intracavity field.

\section{Results and Discussion}

In what follows, we study the radiance witness $R$ of Eq. \eqref{eq:superradiance-witness} for the setup of Fig. \ref{fig:cavity_setup}(a) in a very broad regime of parameters. In Fig. \ref{fig:R_outofphase}, for example, we plot $R$ as a function of the interatomic phase $\phi_z$ and the pumping rate $\eta$ for weak and strong values with respect to the atomic spontaneous emission rate $\gamma$. In all figures we set $\gamma=\kappa$, where $\kappa$ is the cavity decay rate, while the other parameters are varied from figure to figure.
We categorize the value range of $R$ into six different classes, which are depicted in unified colors: extremely subradiant ($R<-0.5$, black), subradiant ($-0.5<R<0$, blue), uncorrelated ($R=0$, light blue), enhanced ($0<R<1$, yellow), superradiant ($R=1$, orange) and hyperradiant ($1<R$, red) scattering. For the color scheme see also the color palette of Fig. \ref{fig:R_outofphase}. Dotted, dashed, and solid curve in the figures indicate a mean photon number $\braket{a^\dagger a}_2=0.01,0.1,1$, respectively. 

\begin{figure}
	\centering \includegraphics{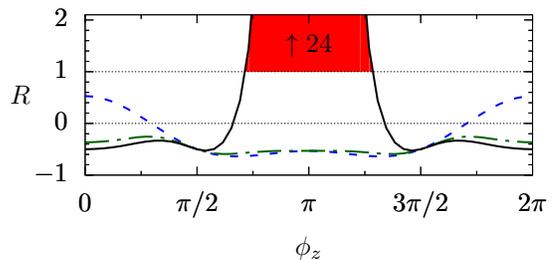}
	\caption{\label{fig:R_outofphase_profiles} Results for different cavities. Vertical cuts of the radiance witness at $\eta\approx0.5\kappa$ as a function of the interatomic phase $\phi_z$ for different types of cavities: blue (dashed) for a bad cavity corresponding to Fig. \ref{fig:R_outofphase}(b); green (dotdashed) for an intermediate cavity corresponding to Fig. \ref{fig:R_outofphase}(c); and black (bold) with highlighted hyperradiant area ($R>1$) for a good cavity corresponding to Fig. \ref{fig:R_outofphase}(a).}
\end{figure}

In good cavities and for atoms radiating out of phase the system can exhibit the phenomenon of hyperradiance, see Fig. \ref{fig:R_outofphase}(a). The radiation can exceed the one of two atoms emitting in phase with otherwise identical parameters distinctly, thereby also surpassing the free-space limit $R=1$. 
This is the synergy of two effects:
Higher-order processes can populate the doubly excited atomic Dicke state $\ket{ee}$, see Fig. \ref{fig:cavity_setup}(b) and (c). In the case of $\phi_z=\pi$, this leads to the emission of single photons into the cavity via the transition $\ket{ee,n}\overset{\gamma}{\rightarrow} \ket{-,n} \overset{g_{-}}{\rightarrow} \ket{gg,n+1}$ or even photon pairs via $\ket{ee,n}\overset{g_{-}}{\rightarrow} \ket{-,n+1} \overset{g_{-}}{\rightarrow} \ket{gg,n+2}$ producing superradiant or even hyperradiant light.
For $\phi_z=0$, however, cavity backaction prevents the excitation of the atoms. This is due to vacuum Rabi splittings \cite{Agarwal:1984,Zhu:1990,Thompson:1992,Tabuchi:2014,Abdurakhimov:2015} of the intracavity field, which for a driving laser on resonance leads to a suppressed excitation of the atoms. The latter can also be interpreted as a destructive quantum path interference \cite{Fernandez-Vidal:2007} between the laser-induced excitation $\ket{gg,n}\overset{\eta}{\rightarrow}\ket{+,n}$ and the cavity-induced excitation $\ket{gg,n+1}\overset{g_{+}}{\rightarrow}\ket{+,n}$, see Fig. \ref{fig:cavity_setup}(b), resulting in subradiant light. The interpretation holds true for uncorrelated atoms, where the interfering terms can be seen in Fig. \ref{fig:cavity_setup}(d) and read $\ket{g,n}\overset{\eta}{\rightarrow}\ket{e,n}$ and $\ket{g,n+1}\overset{g}{\rightarrow}\ket{e,n}$, respectively, which when superimposed yield little excitation of the atom. For two atoms radiating out of phase, however, this back reaction is suppressed as the cavity couples to the anti-symmetric Dicke state $\ket{-}$ and thus the pathway $\ket{gg,n+1}$ to $\ket{+,n}$ is not allowed. As a result, we observe hyperradiance. 

Note that in contrast to the coherent light emitted by a laser, the hyperradiant light is (super-)bunched (as revealed by a second-order correlation function at zero time $g^{(2)}(0) > 1$) due to the emission of photon pairs in the out-of-phase configuration (see Fig. \ref{fig:cavity_setup}(c)). Moreover, commonly lasing is observed when atoms radiate in phase \cite{Meiser:2009}. Opposed to that, in the investigated system the atoms radiate out of phase in the hyperradiant regime.

\begin{figure}
	\centering \includegraphics{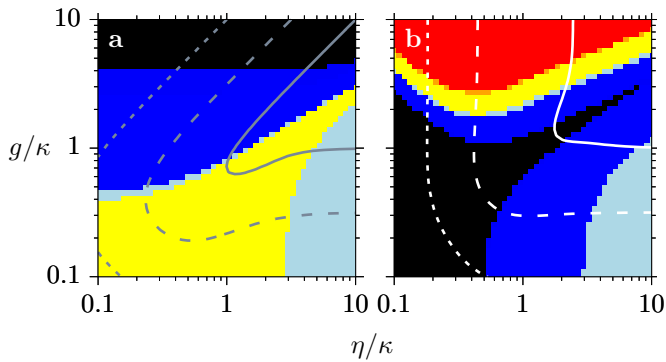}
	\caption{\label{fig:R_inphase} Comparison of in-phase and out-of-phase radiation. Both figures constitute a plot of $R$ as a function of pumping rate $\eta$ and atom-cavity coupling $g$, where $g=0.1\kappa \rightarrow 10\kappa$ reflects the transition from bad to good cavities. Results are shown for $\gamma=\kappa$, no detuning and (a) atoms radiating in phase ($\phi_z=0$), (b) atoms radiating out of phase ($\phi_z=\pi$). For the clarification of the color code as well as dotted, dashed and solid line, see Fig. \ref{fig:R_outofphase}.}
\end{figure}

In intermediate and bad cavities with $g \lesssim \kappa$ the radiation of two atoms out of phase is, however, highly suppressed, see Fig. \ref{fig:R_outofphase}(b) and (c). At $\phi_z=\pi$, the cavity couples to the anti-symmetric Dicke state $\ket{-}$, which is often also called the dark state \cite{Casabone:2015}. When the atoms are driven well below saturation, the coherent laser only pumps the symmetric Dicke state $\ket{+}$ (bright state). 
As a consequence, the cavity mode is almost empty due to destructive interference of the radiation emitted from the two atoms \cite{Fernandez-Vidal:2007}. The radiant character is extremely subradiant $R<-0.9$ or $\braket{a^\dagger a}_2 < 0.2 \braket{a^\dagger a}_1$. 
By contrast, two atoms emitting in phase into an intermediate cavity change their radiant character at higher pumping $\eta$. Here, even at low pumping rates photons can be emitted into the cavity mode via the laser-pumped state $\ket{+}$. 
In Fig. \ref{fig:R_outofphase}(c), for instance, for $\eta \lesssim \kappa$ the pumping strength is not sufficient to pump both atoms leading to a subradiant behavior. At higher $\eta$, both emitters at first scatter uncorrelated, while then higher order processes via $\ket{ee}$ reinforce the atom-cavity coupling leading to an enhanced radiation. If $\eta$ gets too high, the already mentioned destructive quantum path interference takes place. This also occurs at high pumping rates in bad cavities, see Fig. \ref{fig:R_outofphase}(b), while at lower $\eta$ the in-phase radiation is mainly enhanced. In the limit of an extremely bad cavity, corresponding to a free-space setup, superradiant scattering is recovered ($R \rightarrow 1$) for an in-phase configuration.

The detuning can change the radiant behavior drastically, see Fig. \ref{fig:R_outofphase}(d) and (e). By comparing to the undetuned results of Fig. \ref{fig:R_outofphase}(a), we can infer that small detuning of the order of $\delta=\Delta=\kappa$ (d) weakens the hyperradiant behavior while in systems with stronger detuning of the order of $\delta=\Delta=10\kappa$ (e), no hyperradiance can be observed and the light is predominantly subradiant for atoms radiating out-of-phase.

In the experimental realization by Reimann et al. \cite{Reimann:2015}, the authors measure the intensity of the system in a regime where the radiation is suppressed independent of the interatomic phase. Using the parameters of \cite{Reimann:2015}, we observe a transition of the witness from $R=-0.37$ in case of $\phi_z=0$ to $R=-1.00$ in case of $\phi_z=\pi$. Thus, the system becomes extremely subradiant, as the atoms tend to radiate out-of-phase. 
In fact, one could guess that atoms radiating out-of-phase scatter predominantly subradiantly, as observed in all previously mentioned experiments \cite{Reimann:2015,Casabone:2015,Neuzner:2016}. This is the case in bad and intermediate cavities (see dashed and dot-dashed curve in Fig. \ref{fig:R_outofphase_profiles}), or at high detuning. Yet, when studying the behavior in a good cavity and zero detuning, we find that the number of photons within the system can become much larger than in the corresponding setup with uncorrelated atoms. The bold line of Fig. \ref{fig:R_outofphase_profiles} displays this tendency of $R$ for $\eta \approx 0.5\kappa$. Here, the transition of atoms radiating in phase to atoms radiating out of phase is accompanied by the transition from (extreme) subradiance to hyperradiance. In order to observe hyperradiance, the previous experiments \cite{Reimann:2015,Casabone:2015,Neuzner:2016} would need to adapt to the parameters of Figs. \ref{fig:R_outofphase}(a) and \ref{fig:R_inphase}(b).

A brief comparison of atoms located at anti-nodes of a cavity can be seen in Fig. \ref{fig:R_inphase}. Here, the radiation of two atoms radiating in-phase ($\phi_z=0$) and two atoms radiating out-of-phase ($\phi_z=\pi$) is compared over a wide range of coupling constants $g: 0.1\kappa \rightarrow 10\kappa$ reflecting the transition from a bad to a good cavity.  
%Here, we consider this setup for atoms with $\gamma=\kappa$ radiating (a) in-phase and (b) out-of-phase. 
For an in-phase radiation of the atoms, see Fig. \ref{fig:R_inphase}(a), the radiant character hardly depends on the pumping rate as long as $\eta \lesssim \kappa$ but is determined by $g$: for $g\gtrsim 0.5\kappa$ ($g\lesssim 0.5\kappa$), the radiation is subradiant (enhanced) and at $g\approx 0.5\kappa$ uncorrelated. 
Note that $g/\kappa \rightarrow 0$ reflects a free-space setting for which our calculations show that superradiant scattering is recovered, $R\rightarrow 1$.
%At $\eta>\kappa$, the enhanced radiation part increases, while at very high $\eta$, destructive interference arises.
In Fig. \ref{fig:R_inphase}(b), we compare these findings to atoms radiating out of phase in the same parameter range. While for atoms radiating in phase, the transition from bad to good cavities goes along with the transfer from superradiance or enhanced radiation to subradiance, the situation is reversed for atoms radiating out of phase. Here they radiate subradiantly in bad cavities, whereas their radiation in good cavities can exceed the superradiant limit distinctly, finally ending up in hyperradiance, which can be explained via quantum path interference.

\begin{figure}
	\centering 
	\includegraphics{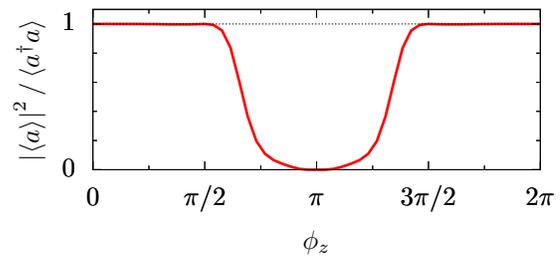}
	\caption{Comparison of classical and quantum mechanical treatment. The ratio $\left|\braket{a}\right|^2/\braket{a^\dagger a}$ comparing the classical intensity of the intracavity field with the quantum-mechanical mean photon number is shown as a function of the interatomic phase $\phi_z$ for $g=10\kappa$, $\gamma=\kappa$ and $\eta=0.1\kappa$.}
	\label{fig:quantum}
\end{figure}

Interestingly, the classical treatment of the discussed setup predicts an intracavity field that vanishes in the case of an out-of-phase configuration, see Eq. \eqref{eq:classical-field} with $\mathcal{G}(\phi_z=\pi)=0$. One can quantify the deviation from the classical approach by considering the ratio $\left|\braket{a}\right|^2/\braket{a^\dagger a}$, which compares the classical intensity of the intracavity field with the quantum-mechanical mean photon number. A deviation from unity reveals quantum features displayed by the system. For the investigated two atom-cavity system, $\left|\braket{a}\right|^2/\braket{a^\dagger a}$ equals one for an in-phase configuration, but tends to zero as $\phi_z\rightarrow \pi$, see Fig. \ref{fig:quantum}.
A value below one of the ratio displayed in Fig. \ref{fig:quantum} corresponds to the quantum theory predicting a higher intensity than the classical approach.
Therefore, the occurrence of hyperradiance even in the low pumping regime $\eta \approx 0.1 \kappa$ can only be explained in a full quantum-mechanical treatment revealing the true quantum origin of the phenomenon hyperradiance.

\section{Conclusions}

In conclusion, we have shown a new phenomenon in the collective behavior of coherently driven atoms, which we call hyperradiance. In this regime, the radiation of two atoms in a single-mode cavity coherently driven by an external laser can exceed the free-space superradiant behavior considerably. Hyperradiance occurs in good cavities and, surprisingly, for atoms radiating out of phase. The effect cannot be explained in a \mbox{(semi-)}classical treatment revealing a true quantum origin. Moreover, by modifying merely the interatomic phase, crossovers from subradiance to hyperradiance can be observed. 
Our results should stimulate various new experiments examining the possibility for the observation of hyperradiance in this fundamental system, consisting of a cavity coupled to any kind of two-level systems like atoms, ions, superconducting qubits or quantum dots. 

\section*{Acknowledgments}

M.-O.P. gratefully acknowledges the hospitality at the Oklahoma State University. 
The authors gratefully acknowledge funding by the Erlangen Graduate School in Advanced Optical Technologies (SAOT) by the German Research Foundation (DFG) in the framework of the German excellence initiative. 
Some of the computing for this project was performed at the OSU High Performance Computing Center at Oklahoma State University supported in part through the National Science Foundation grant OCI–1126330. 

\bibliography{bib}

\end{document}